# Nanosecond spin relaxation times in single layer graphene spin valves with hexagonal boron nitride tunnel barriers

Simranjeet Singh[1], Jyoti Katoch[1], Jinsong Xu[1], Cheng Tan[2], Tiancong Zhu[1], Walid Amamou[3], James Hone[2] and Roland Kawakami[1,3*]

[1]Department of Physics, The Ohio State University, Columbus, Ohio, United States 43210

[2]Mechanical Engineering Department, Columbia University, New York, NY, United States 10027

[3]Program of Materials Science and Engineering, University of California, Riverside, CA, United States 92521

**Abstract:**

We present an experimental study of spin transport in single layer graphene using atomic sheets of hexagonal boron nitride (h-BN) as a tunnel barrier for spin injection. While h-BN is expected to be favorable for spin injection, previous experimental studies have been unable to achieve spin relaxation times in the nanosecond regime, suggesting potential problems originating from the contacts. Here, we investigate spin relaxation in graphene spin valves with h-BN barriers and observe room temperature spin lifetimes in excess of a nanosecond, which provides experimental confirmation that h-BN is indeed a good barrier material for spin injection into graphene. By carrying out measurements with different thicknesses of h-BN, we show that few layer h-BN is a better choice than monolayer for achieving high non-local spin signals and longer spin relaxation times in graphene.

[*] Author to whom correspondence should be addressed: kawakami.15@osu.edu

Graphene is a promising spin channel material for next generation spintronic devices due to the experimental demonstration of long spin diffusion lengths at room temperature[1-3] and theoretical predictions of long spin relaxation times[4,5] arising from the weak spin-orbit and hyperfine couplings[5,6]. However, experimentally measured spin relaxation times[1-3,7,8] in graphene are orders of magnitude shorter than theoretically predicted[4,5]. In graphene spin valves, the tunnel barrier plays a crucial role for spin injection by circumventing the problem of impedance mismatch[9] between graphene and the ferromagnetic electrodes. As demonstrated by Han et. al.[8], high quality tunnel barriers are critical for obtaining higher spin relaxation times ($\tau_s$) in graphene because barriers with pinholes or rough surface morphology can cause additional contact-induced spin relaxation, which has received a great deal of interest recently.[10-14] As opposed to growing oxide tunnel barriers on graphene, a thin insulating two-dimensional (2D) van der Waals material can also be used as a tunnel barrier. A particular material of interest is single (or few) layer h-BN because of its various suitable properties[15]: large energy band gap ~5.97 eV, high crystallinity, spin filtering[16], absence of pinholes and dangling bonds, atomic lattice similar to graphene, and chemical stability at ambient conditions. In addition, atomically clean vertical heterostructures of h-BN/graphene can be mechanically assembled using polymer-based transfer techniques[17,18]. The first experimental report demonstrating spin injection into graphene using a monolayer h-BN tunnel barrier showed $\tau_s$ less than 100 ps[19]. This was followed by the work of Kamalakar et. al.[20,21] and Fu et. al.[22], which used chemically grown h-BN barriers, yielding $\tau_s$ ~ 500 ps. Another recent study using an encapsulated geometry[23] with graphene sandwiched between a thick bottom layer of h-BN and a monolayer of h-BN on top showed $\tau_s$ less than 200 ps. As evident from these studies, graphene spin valve devices with h-BN tunnel barriers have yielded relatively small values for $\tau_s$ (< 1 ns) that are comparable to or lower than the values obtained using oxide barriers. Thus, it is worthwhile to ask: are h-BN tunnel barriers compatible with longer spin lifetimes in graphene?

In this letter, we perform spin transport in single layer graphene spin valve devices with h-BN tunnel barriers and observe spin relaxation times exceeding a nanosecond at room temperature, the highest values achieved so far for devices employing h-BN tunnel barriers. In addition, we investigate the thickness dependent characteristics of the h-BN tunnel barriers and find that few layer h-BN, rather than monolayer, is required to observe large non-local spin signals and longer spin relaxation times in graphene. Our work establishes the effectiveness of ultrathin h-BN as a high quality tunnel barrier for spin injection into graphene, which is a crucial step towards realizing high performance spintronic devices based completely on ultrathin van der Waals materials.

For making h-BN/graphene interfaces, thin h-BN flakes are exfoliated from bulk crystals (HQ graphene) on a 90 nm $SiO_2$/Si wafer, and the thickness of h-BN is confirmed by atomic force microscopy (AFM). On a separate 300 nm $SiO_2$/Si substrate, Kish graphite is exfoliated to obtain single layer graphene, confirmed by Raman spectroscopy. Figure 1a shows the Raman spectra of the graphene used in the h-BN/graphene heterostructures. Thin h-BN is transferred onto single layer graphene using a process similar to Zomer et. al.[24] In short, a polydimethylsiloxane (PDMS) stamp coated with polycarbonate (PC) polymer is attached to a glass slide and is brought in contact with an exfoliated h-BN flake. Next, the PC film is heated to 70 °C to pick-up the h-BN flake from the $SiO_2$ surface. Once the h-BN is on the PC film, it is optically aligned to the desired graphene flake and then the PC film is melted onto the graphene flake's substrate at 150 °C. The PC film is removed by dissolving it in chloroform for 30 minutes. Afterward, the h-BN/graphene heterostructure is cleaned by annealing in $H_2$/Ar forming gas at 350 °C for 3 hours. Figure 1c shows the surface topography of the measured h-BN/graphene stack, where the top h-BN and the graphene are outlined by yellow and red dashed lines, respectively. The measured thickness of the h-BN is 0.85 nm, as shown by the step height in the inset of Figure 1b, which corresponds to 2-3 layers of h-BN[25] (monolayer thickness is ~0.34 nm). Although there are few small bubbles trapped between the graphene and h-BN, there are still large bubble-free regions that can be used for the cobalt contacts. Figure 1c shows the optical microscope image of the measured spin valve device, in which yellow dotted lines highlight the boundary of the h-BN tunnel barrier. In order to electrically characterize the graphene sheet, we measure the four-probe resistance (at room temperature) of graphene as a function of back gate voltage ($V_G$) applied to the degenerately doped Si substrate (Figure 1d). An electron mobility of 4000 cm$^2$/Vs is extracted from the slope of the graphene conductivity versus gate voltage scan[26]. The contact resistance of the electrodes with h-BN barrier is measured using a three-probe measurement configuration[8]. The zero bias contact resistance of the electrodes varies from a few kilo-Ohms (kΩ) to hundreds of kΩ depending on the thickness of the h-BN. This will be discussed in detail later.

We begin by discussing the spin transport measurements carried out at room temperature. For the data presented here, the graphene channel length is 1.5 microns (μm) and width is 2.5 μm. The contact resistances of the injector (E2; Fig. 1c) and detector (E3; Fig. 1c) electrodes are 42 kΩ and 6 kΩ, respectively. The temperature dependence and bias dependence of the contact resistances indicate tunneling behavior (see supplementary material). To measure the non-local magnetoresistance (MR) signal, we perform low frequency ac measurements using a current excitation of 1 μA rms. As shown by the schematic in the inset to Figure 2a, spin injection current is applied between the electrodes labeled E1 and E2, and a non-local voltage signal is measured using the electrodes labeled E3 and E4. An external magnetic field is swept in the plane of the graphene device (along the electrode's length), and a non-local voltage signal is recorded as a function of the applied magnetic field. To obtain the non-local resistance ($R_{NL}$), the voltage signal is divided by the magnitude of applied current (1 μA). Figure 2a shows the observed non-local MR signal at room temperature, where black circles are data recorded by a sweep in increasing magnetic field. At ~20 mT, we see an abrupt change in $R_{NL}$ when one of the magnetizations reverses to create an antiparallel alignment of the injector and detector magnetizations (arrows indicate the relative alignment of the magnetizations throughout the magnetic field sweeps).

This is the hallmark of spin transport from injector to detector. With further increase of the magnetic field, the resistance signal again changes back to the original value due to the alignment of the injector and detector magnetizations back to a parallel configuration. Also shown in Figure 2a is the MR signal when sweeping with a decreasing magnetic field (blue circles). The magnitude of the measured MR signal ($\Delta R_{NL}$) is defined as the difference of $R_{NL}$ between the parallel and antiparallel configurations, and is approximately 4 $\Omega$ for the measured device at $V_G$ = +10 V. We also measure the gate dependence of the non-local MR on the applied gate voltage, which tunes the polarity and density of the charge carriers and observe $\Delta R_{NL}$ as large as 5 $\Omega$ over the measured gate voltage range (data shown in the supplementary Figure S2b).

To study the spin relaxation times in the graphene channel, we perform non-local Hanle spin precession measurements at room temperature. In a typical Hanle measurement, the in-plane polarized spins precess in the graphene plane by an externally applied out of plane magnetic field. After aligning the magnetization of the injector and detector, either in a parallel or antiparallel configuration, a magnetic field is swept perpendicular to the plane of the graphene device to measure the non-local resistance signal as a function of applied magnetic field. Figure 2b shows the Hanle curves obtained for the graphene spin valve device measured in parallel (blue circles) and antiparallel (black circles) configuration. The measured sweeps for parallel and antiparallel configurations are subtracted to obtain the curve as shown in Figure 2c (black circles). In general, once the spins are injected from the ferromagnetic electrode, the spin transport in graphene can be described by majority and minority spin channels. The voltage signal measured at the detector electrode is proportional to net spin accumulation ($\boldsymbol{\mu_S}$), where $|\mu_S|$ is the difference of the electrochemical potentials of majority and minority spins and the vector points along the polarization axis. This spin accumulation in an applied magnetic field can be described by solving the steady state Bloch equation:

$$D\nabla^2 \boldsymbol{\mu}_s - \frac{\boldsymbol{\mu}_s}{\tau_s} + \boldsymbol{\omega}_L \times \boldsymbol{\mu}_s = 0, \tag{1}$$

where $D$ is the diffusion constant in the graphene channel, $\tau_s$ is the spin relaxation time, $\boldsymbol{\omega}_L = g\mu_B \mathbf{B}/\hbar$ is the Larmor frequency, $g$ = 2 is the gyromagnetic factor, $\mu_B$ is the Bohr magneton, $\hbar$ is the reduced Planck's constant, and $\mathbf{B}$ is the externally applied magnetic field. To fit the measured non-local Hanle data we use an analytical expression developed by Sosenko et. al.[27] which is the solution to the steady state Bloch equation (eq. 1) in the presence of boundary conditions for the injection and absorption of spin current at the ferromagnetic electrodes:

$$R_{NL}^{\pm} = \pm p_1 p_2 R_N f \tag{2}$$

where the $\pm$ corresponds to the relative alignment of injector/detector magnetizations, $p_1$ and $p_2$ are the electrode spin polarizations, $R_N = \frac{\lambda}{W L \sigma^N}$ is the spin resistance of graphene, $\lambda$ is the spin diffusion length in graphene (related to $\tau_s$ by $\lambda = \sqrt{D\tau_s}$), $W$ is the width of the graphene, $\sigma^N$ is the conductivity of graphene and

$$f = Re\left( 2\left[\sqrt{1+i\omega_L \tau_s} + \frac{\lambda}{2}\left(\frac{1}{r_s} + \frac{1}{r_d}\right)\right] e^{\left(\frac{L}{\lambda}\right)\sqrt{1+i\omega_L \tau_s}} + \frac{\lambda^2}{r_s r_d} \frac{\sinh\left[\left(\frac{L}{\lambda}\right)\sqrt{1+i\omega_L \tau_s}\right]}{\sqrt{1+i\omega_L \tau_s}} \right)^{-1} \tag{3}$$

In the function $f$, $r_i = \frac{R_F + R_C^i}{R_{SQ}} W$ with $i = s, d$ for the injector and detector, respectively, $R_C$ is the contact resistance, $R_{SQ}$ is the graphene sheet resistance, $R_F = \rho_F \lambda_F / A$ is the spin resistance of the ferromagnetic electrode material, $\rho_F$ is the resistivity of the ferromagnet, $\lambda_F$ is the spin diffusion length in the ferromagnet, and $A$ is the area of the graphene/electrode junction.

Using Equation 2, the experimentally obtained Hanle curves can be fit using $\tau_s$, $\lambda$, and the product $p_1 p_2$ as fitting parameters. The fit for the Hanle curve at zero gate voltage (solid red line in Figure 2c) yields values of $\tau_s$ = 1.86 ns, $\lambda$ = 5.78 µm, $\sqrt{p_1 p_2}$ = 0.053, and $D$ = 0.018 m²/s. The corresponding spin injection efficiency is found to be 0.052 (see supplementary material). We also investigated the gate dependence of the spin lifetimes by measuring Hanle precession curves at different $V_G$ and fitting each of these curves with Equation 2. Figure 2d shows the extracted $\tau_s$ as a function of applied gate voltages, where, for most of the applied gate voltages, $\tau_s$ exceeds 1 ns. This experimental demonstration of nanosecond spin lifetimes in graphene on $SiO_2$ substrates at room temperature, employing h-BN tunnel barriers, is the central result of this work. These observed nanosecond spin lifetimes are the highest values reported in the literature for a single layer graphene channel employing an h-BN tunnel barrier[19-23]. It is important to note that for van der Waals heterostructures, it is non-trivial to achieve clean interfaces[17,28], and the impurities (or residues) at the interface can affect the electrical and spin related properties across the van der Waals heterostructures. One possible explanation for the high quality spin transport observed in our studies may be the relatively clean h-BN/graphene interface, evidenced by the flat surface topography of the heterostructure, in areas where we deposited ferromagnetic electrodes for spin injection.

We have also spin transport in graphene spin valves using monolayer h-BN[19,25] (~0.50 nm thick) tunnel barriers, in contrast to the few layer (2-3) h-BN barriers discussed so far. Here we present data using a pair of electrodes with injector and detector contact resistance of 7.7 kΩ and 3.7 kΩ respectively. The detailed temperature dependence of the contact resistance is shown in supplementary material. The graphene channel length is 5 μm and width is 4.9 μm. The charge carrier mobility of the graphene channel is 8000 cm$^2$/Vs. As shown in Figure 3c, the non-local MR signal measured at 11 K and $V_G$ = +10 V exhibits a magnitude of 170 mΩ. We present data taken at low temperatures because we are unable to resolve clear MR switching at RT beyond the noise, but can observe clear MR at low temperatures. The magnitude of the MR ($\Delta R_{NL}$) as a function of gate voltage varies from 100 mΩ to 350 mΩ (data shown in the supplementary material). Figure 3b shows the Hanle curve (black circles) obtained by carrying out non-local Hanle measurements at a $V_G$ = +10 V. By fitting the data with Equation 2 we extract $\tau_s$ = 490 ps, λ=3.8 μm, $\sqrt{p_1 p_2}$ = 0.023, and D = 0.03 m²/s. Figure 3c shows the extracted $\tau_s$ as a function of applied gate voltage, where the inset shows the gate dependent resistance of the measured graphene channel. The spin relaxation times extracted from the Hanle curves range from 300 ps - 600 ps, which are similar to (or slightly higher than) previous studies employing monolayer h-BN tunnel barriers[19,23]. Our measurements suggest that monolayer h-BN may not be the optimal choice for efficient spin injection, while thicker h-BN allows the realization of larger MR signals and longer spin relaxation times in graphene spin valves. One of the possible reasons for higher spin relaxation times in graphene using multilayer h-BN tunnel barriers could be related to the fact that few layer h-BN (on SiO$_2$ substrate) is flatter[17], i.e. less ripples in surface morphology, than single layer h-BN. This could lead to smoother growth of cobalt electrodes on multilayer h-BN and thus reducing the contact-induced spin dephasing mechanism due to the magnetostatic fringe field[29] of the ferromagnetic material. Apart from this, we also speculate that single layer h-BN is more prone to surface contaminants such as organic residues or wrinkles during the transfer process and can cause additional spin dephasing under the ferromagnet contact resulting in shorter spin relaxation times in graphene.

In order to investigate the range of h-BN thickness required to observe large MR signals and higher spin relaxation times in graphene spin valves, we have prepared different h-BN/graphene stacks with h-BN thicknesses varying from ~ 0.50 nm to 3.20 nm. The thicknesses of the h-BN flakes were characterized by AFM. In total, we measured six different devices with h-BN tunnel barriers. For each device with multiple ferromagnetic electrodes, the interfacial contact resistance was measured using a 3-probe configuration[8], wherein the differential contact resistance (dV/dI) was measured as function of dc bias current at low temperature. The dV/dI curves show a zero bias peak, as expected for a tunnel barrier. The values of zero bias dV/dI as function of the h-BN thickness are plotted in Figure 4 for all the measured electrodes. The gray dotted circle denotes the region, corresponding to h-BN thicknesses, for which we were able to observe MR signals. We did not observe any non-local MR signal in devices with h-BN thickness larger than 1.50 nm (~4 monolayers of h-BN). Note that one would expect to have a small distribution of contact resistances for a given h-BN thickness, but the interfacial inhomogeneities (e.g., bubbles or organic residues) at the h-BN/graphene interface over mesoscopic size areas could give rise to the observed contact resistance distributions.

In conclusion, our experimental work demonstrates that h-BN is a high quality tunnel barrier material for spin injection into graphene, as evident from our data showing large non-local MR signals and nanosecond spin relaxation times at room temperature. Our experimental observations indicate that few layer h-BN leads in better characteristics, as opposed to monolayer h-BN, for tunnel barrier applications to achieve higher quality spin transport in graphene. More experimental studies are needed to further improve the interface quality of h-BN/graphene in order to exploit the full potential of van der Waals heterostructures for spin related physics in graphene and other 2D materials.

**Acknowledgements:**

The authors acknowledge Elizabeth Bushong for carefully reading the manuscript. S.S., J.K., J.X., T.Z., W.A. and R.K. acknowledge support from ONR (No. N00014-14-1-0350), NRI-NSF (No. DMR-1124601), NSF (No. DMR-1310661), and C-SPIN, one of the six SRC STARnet Centers, sponsored by MARCO and DARPA. C.T was supported by a DOD-AFOSR, NDSEG fellowship under contract FA9550-11-C-0028, 32 CFR 168a. C.T. and J.H. acknowledge support from the Nanoelectronics Research Initiative (NRI) through the Institute for Nanoelectronics Discovery and Exploration (INDEX).

See supplementary material for detailed discussion and data for: temperature dependence of the interfacial contact resistances, gate dependent MR signals and h-BN thickness dependence of the spin injection efficiency.

# References:

1. M. H. D. Guimarães, P. J. Zomer, J. Ingla-Aynés, J. C. Brant, N. Tombros, and B. J. van Wees, Physical Review Letters **113**, 086602 (2014).
2. M. Drögeler, F. Volmer, M. Wolter, B. Terrés, K. Watanabe, T. Taniguchi, G. Güntherodt, C. Stampfer, and B. Beschoten, Nano Letters **14**, 6050 (2014).
3. M. Drögeler, C. Franzen, F. Volmer, T. Pohlmann, L. Banszerus, M. Wolter, K. Watanabe, T. Taniguchi, C. Stampfer, and B. Beschoten, Nano Letters **16**, 3533 (2016).
4. D. Huertas-Hernando, F. Guinea, and A. Brataas, Physical Review B **74**, 155426 (2006).
5. W. Han, R. K. Kawakami, M. Gmitra, and J. Fabian, Nat Nano **9**, 794 (2014).
6. A. H. Castro Neto, F. Guinea, N. M. R. Peres, K. S. Novoselov, and A. K. Geim, Reviews of Modern Physics **81**, 109 (2009).
7. N. Tombros, C. Jozsa, M. Popinciuc, H. T. Jonkman, and B. J. van Wees, Nature **448**, 571 (2007).
8. W. Han, K. Pi, K. M. McCreary, Y. Li, J. J. I. Wong, A. G. Swartz, and R. K. Kawakami, Physical Review Letters **105**, 167202 (2010).
9. G. Schmidt, D. Ferrand, L. W. Molenkamp, A. T. Filip, and B. J. van Wees, Physical Review B **62**, R4790 (2000).
10. M. Popinciuc, C. Józsa, P. J. Zomer, N. Tombros, A. Veligura, H. T. Jonkman, and B. J. van Wees, Physical Review B **80**, 214427 (2009).
11. T. Maassen, I. J. Vera-Marun, M. H. D. Guimarães, and B. J. van Wees, Physical Review B **86**, 235408 (2012).
12. W. Amamou, Z. Lin, J. van Baren, S. Turkyilmaz, J. Shi, and R. K. Kawakami, APL Mater. **4**, 032503 (2016).
13. H. Idzuchi, A. Fert, and Y. Otani, Physical Review B **91**, 241407 (2015).
14. H. Idzuchi, Y. Fukuma, S. Takahashi, S. Maekawa, and Y. Otani, Physical Review B **89**, 081308 (2014).
15. K. Watanabe, T. Taniguchi, and H. Kanda, Nat Mater **3**, 404 (2004).
16. M. V. Kamalakar, A. Dankert, P. J. Kelly, and S. P. Dash, Scientific Reports **6**, 21168 (2016).
17. C. R. Dean, A. F. Young, MericI, LeeC, WangL, SorgenfreiS, WatanabeK, TaniguchiT, KimP, K. L. Shepard, and HoneJ, Nat Nano **5**, 722 (2010).
18. L. Wang, I. Meric, P. Y. Huang, Q. Gao, Y. Gao, H. Tran, T. Taniguchi, K. Watanabe, L. M. Campos, D. A. Muller, J. Guo, P. Kim, J. Hone, K. L. Shepard, and C. R. Dean, Science **342**, 614 (2013).
19. Y. Takehiro, I. Yoshihisa, M. Satoru, M. Sei, O. Masahiro, W. Kenji, T. Takashi, M. Rai, and M. Tomoki, Applied Physics Express **6**, 073001 (2013).
20. M. V. Kamalakar, A. Dankert, J. Bergsten, T. Ive, and S. P. Dash, Applied Physics Letters **105**, 212405 (2014).
21. M. V. Kamalakar, A. Dankert, J. Bergsten, T. Ive, and S. P. Dash, Scientific Reports **4**, 6146 (2014).
22. W. Fu, P. Makk, R. Maurand, M. Bräuninger, and C. Schönenberger, Journal of Applied Physics **116**, 074306 (2014).
23. M. Gurram, S. Omar, S. Zihlmann, P. Makk, C. Schönenberger, and B. J. van Wees, Physical Review B **93**, 115441 (2016).
24. P. J. Zomer, M. H. D. Guimarães, J. C. Brant, N. Tombros, and B. J. van Wees, Applied Physics Letters **105**, 013101 (2014).
25. G.-H. Lee, Y.-J. Yu, C. Lee, C. Dean, K. L. Shepard, P. Kim, and J. Hone, Applied Physics Letters **99**, 243114 (2011).
26. J.-H. Chen, C. Jang, S. Xiao, M. Ishigami, and M. S. Fuhrer, Nat Nano **3**, 206 (2008).

27. E. Sosenko, H. Wei, and V. Aji, Physical Review B **89**, 245436 (2014).
28. S. J. Haigh, A. Gholinia, R. Jalil, S. Romani, L. Britnell, D. C. Elias, K. S. Novoselov, L. A. Ponomarenko, A. K. Geim, and R. Gorbachev, Nat Mater **11**, 764 (2012).
29. S. P. Dash, S. Sharma, J. C. Le Breton, J. Peiro, H. Jaffrès, J. M. George, A. Lemaître, and R. Jansen, Physical Review B **84**, 054410 (2011).

# Figure Captions:

**Figure 1**: (a) Raman spectra confirming single layer graphene. (b) Atomic force microscopy of the h-BN/graphene stack (before defining the ferromagnetic electrodes) showing the topography of graphene and thin h-BN. The boundaries of the graphene and top h-BN are highlighted by red and yellow dotted line, respectively. The thickness of the h-BN is ~0.85 nm (2-3 monolayers) and the step height of h-BN is shown in the inset. (c) The optical image of the completed device, with different electrodes used for the measurements labeled for convenience. The h-BN flake boundary is depicted by yellow dotted lines. (d) The gate dependent resistance of the graphene channel measured at room temperature.

**Figure 2**: Spin transport in single layer graphene using few layer h-BN tunnel barriers, measured at room temperature. (a) Non-local magneto-resistance (MR) signal measured in a graphene spin valve with h-BN tunnel barriers using E2 and E3 as injector and detector electrodes, respectively, at room temperature. The blue and black arrows represent the relative magnetization direction of injector and detector electrodes. $\Delta R_{NL}$ is the magnitude of the MR signal. Inset: schematic of the non-local measurement configuration. (b) Non-local Hanle spin precession signal measured in parallel (blue circles) and antiparallel (black circles) configuration by applying an out of plane magnetic field. The blue and black arrows represents the relative magnetization direction of injector and detector. (c) Non-local Hanle spin precession curve (black circles) obtained by subtracting parallel and antiparallel Hanle curves from Figure 2b. The thick red line is the fit to the data to extract the spin relaxation time ($\tau_s$). (d) The gate dependence of the fitted spin relaxation times over a range of applied back gate voltages.

**Figure 3**: Spin transport in single layer graphene using monolayer h-BN tunnel barriers, measured at 11 K. (a) Non-local magnetoresistance (MR) signal measured in a graphene spin valve using monolayer h-BN at $V_G$ = +10 V. (b) Non-local Hanle signal (black circles) measured by applying an out of plane magnetic field at $V_G$ = +10 V. The red line is the fitting to the data to extract spin relaxation time ($\tau_s$). (c) The fitted spin relaxation times as function of applied back gate. Inset: graphene channel resistance as a function of gate voltage.

**Figure 4**: Low temperature tunneling characteristics of different thickness h-BN barriers. Differential contact resistance of the electrodes (for 6 different devices) is shown as a function of AFM measured thickness of the h-BN layer. The gray circle shows the region for which we have observed MR signals.

Fig. 1

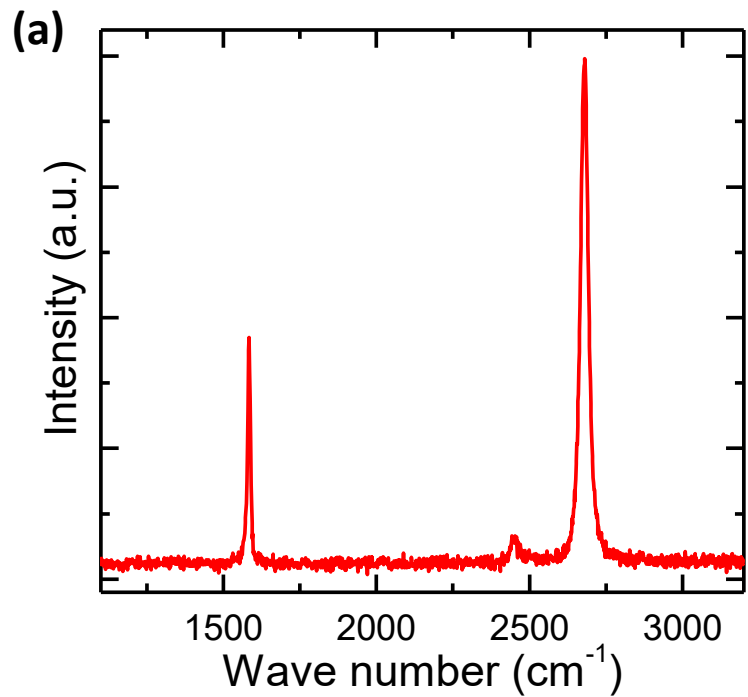
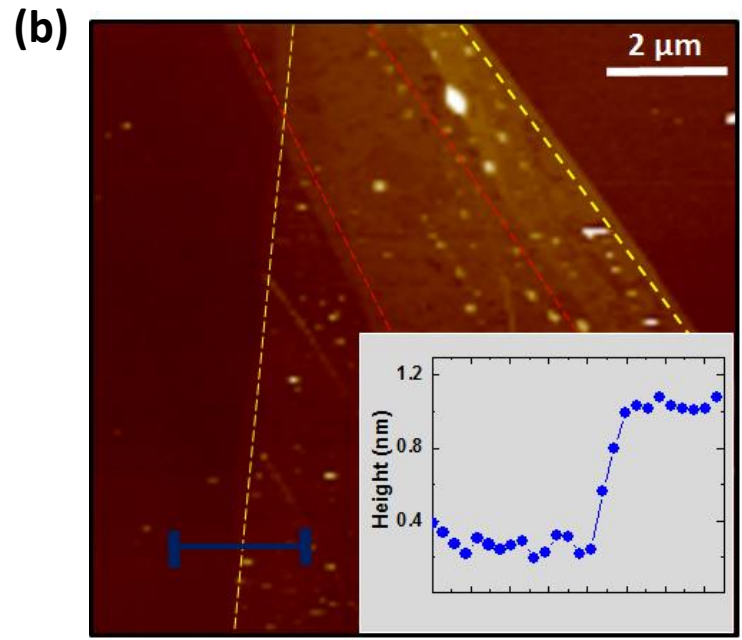
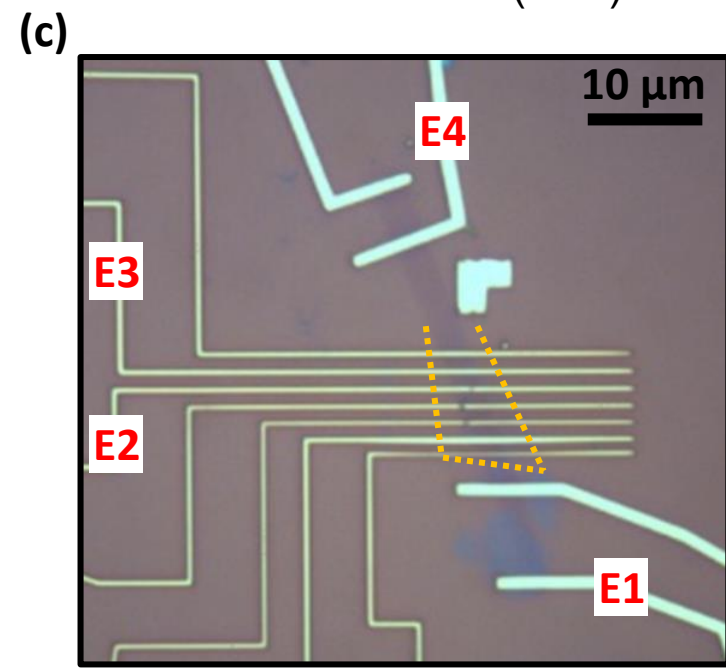
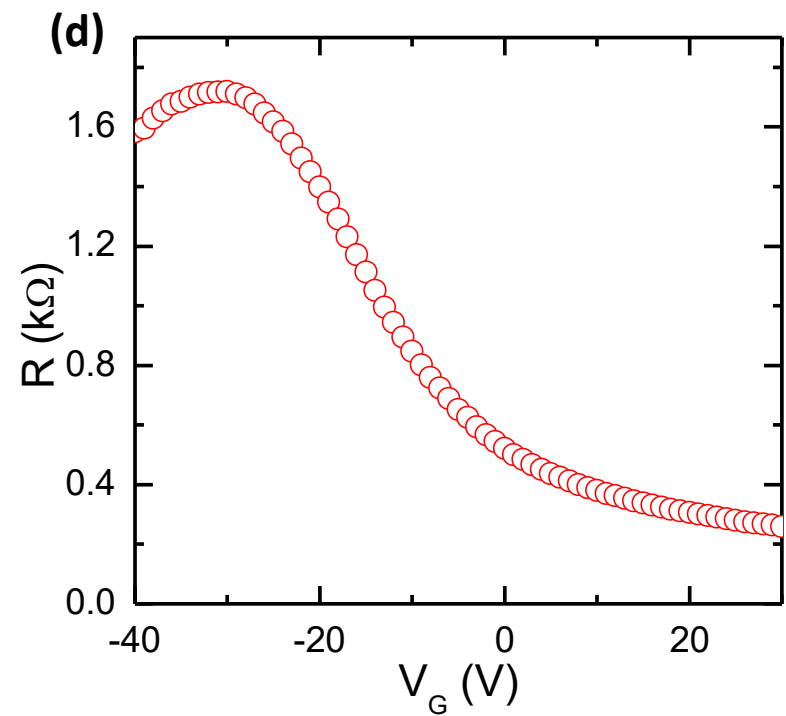

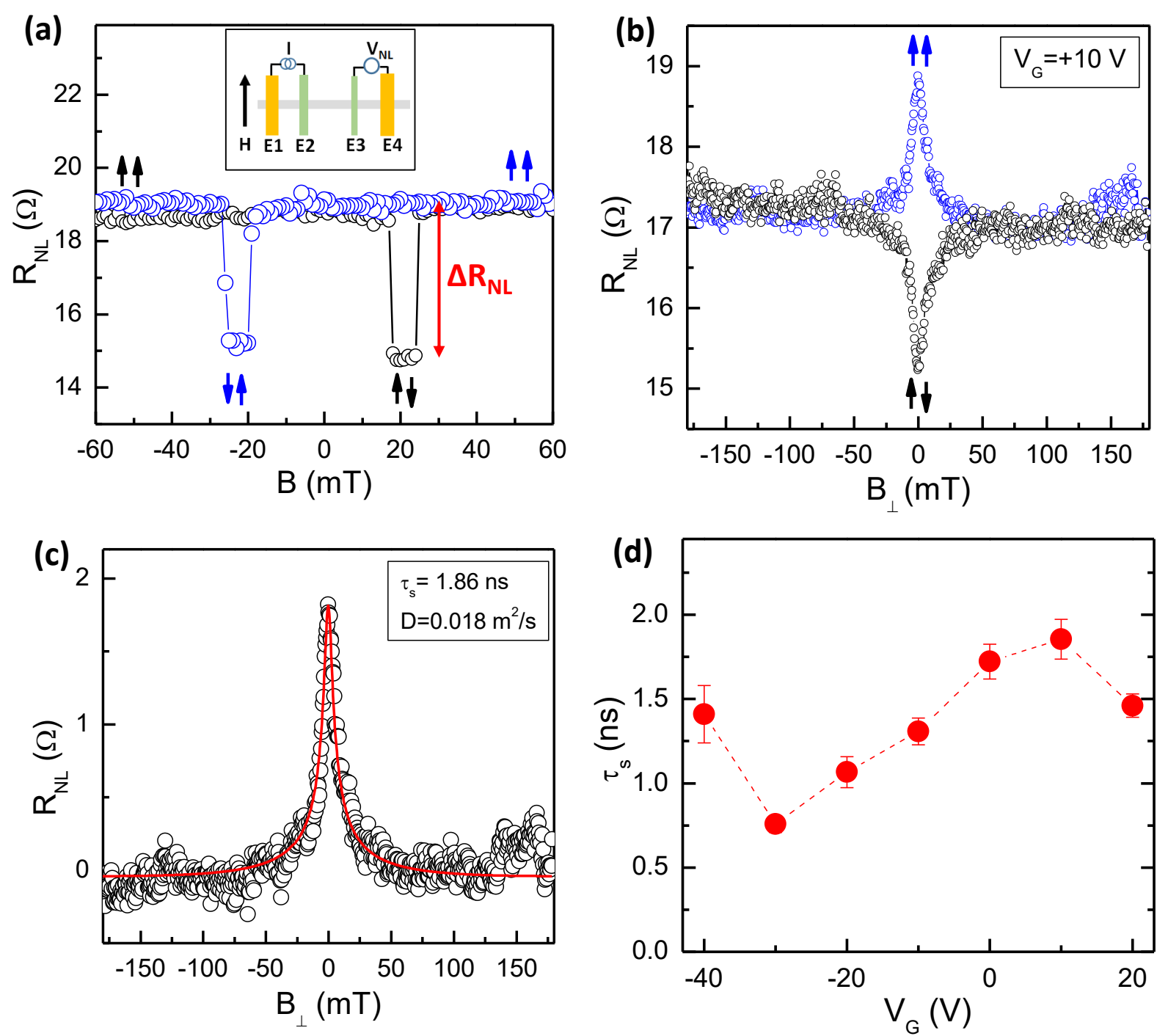

Fig. 2

Fig. 3

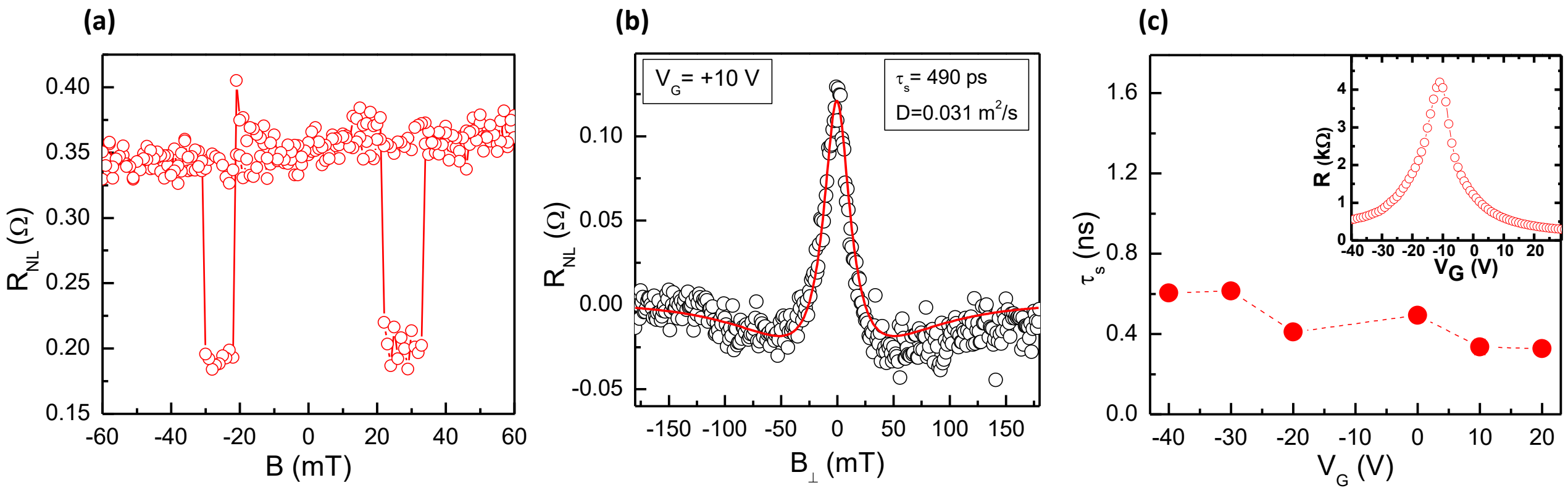

Fig. 4

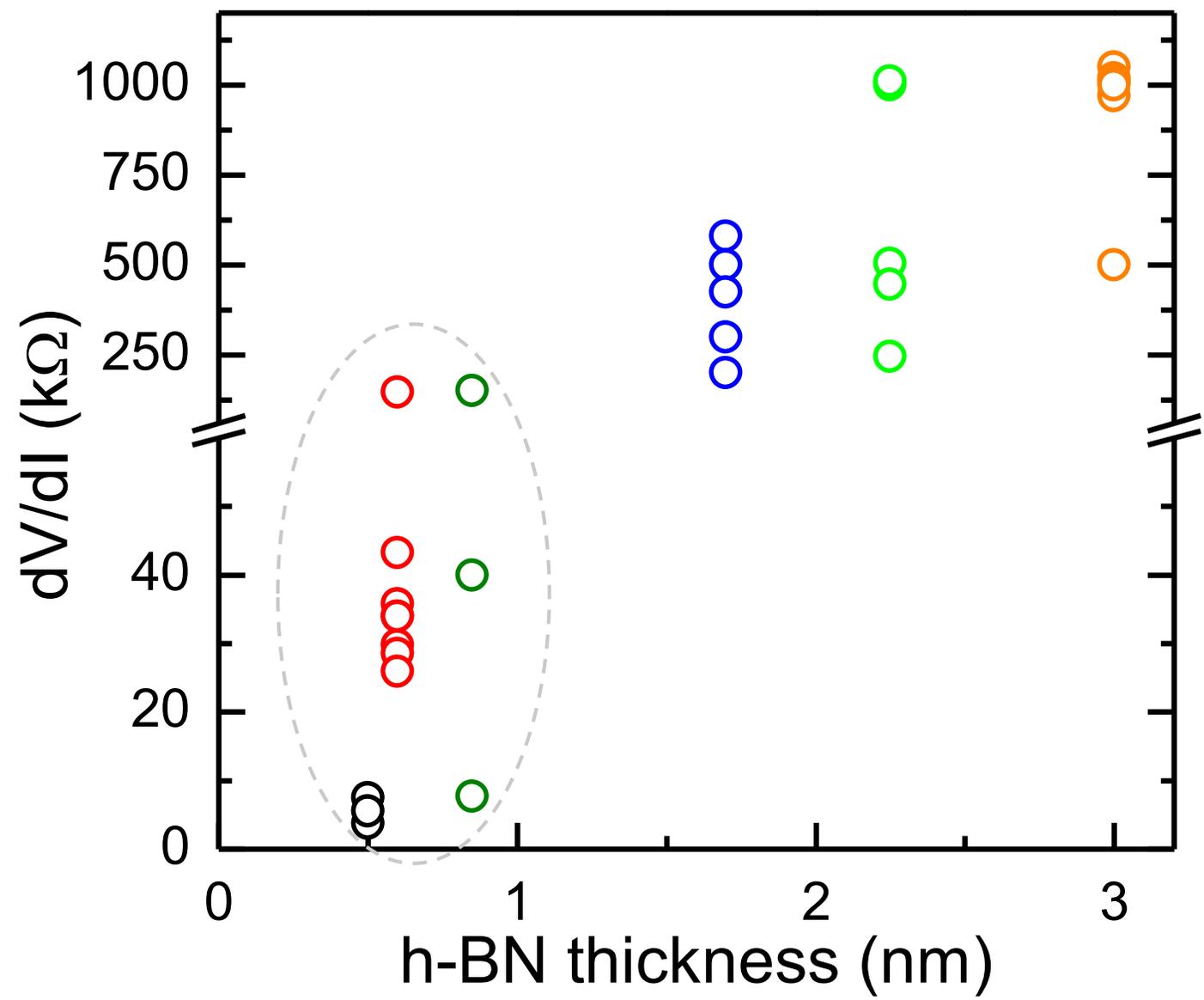

**Supplemental Material**

# Nanosecond spin relaxation times in single layer graphene spin valves with hexagonal boron nitride tunnel barriers

Simranjeet Singh[1], Jyoti Katoch[1], Jinsong Xu[1], Cheng Tan[2], Tiancong Zhu[1], Walid Amamou[3], James Hone[2] and Roland Kawakami[1,3]

[1]Department of Physics, The Ohio State University, Columbus, Ohio, United States 43210

[2]Mechanical Engineering Department, Columbia University, New York, NY, United States 10027

[3]Program of Materials Science and Engineering, University of California, Riverside, CA, United States 92521

## I. Temperature dependence of contact resistances

We present the temperature dependence of the contact resistances of the injector and detector electrodes. For the 0.85 nm thick h-BN tunnel barrier, the inset to the Figure S1a shows the measured dV/dI curve as a function of dc bias current for both the injector and detector electrodes. The zero bias contact resistance of injector and detector electrodes, for which we have presented data in the main text, shows very weak temperature dependence from 11 K to 300 K. This indicates tunneling contacts, in contrast to pinhole barriers which display a sharp increase in contact resistance with increasing temperature. Similarly, the temperature dependence of contact resistance of the 0.50 nm thick h-BN tunnel barrier is shown in Figure S1b. The dc bias dependence of the contact resistance is shown in the inset to the Figure S1b.

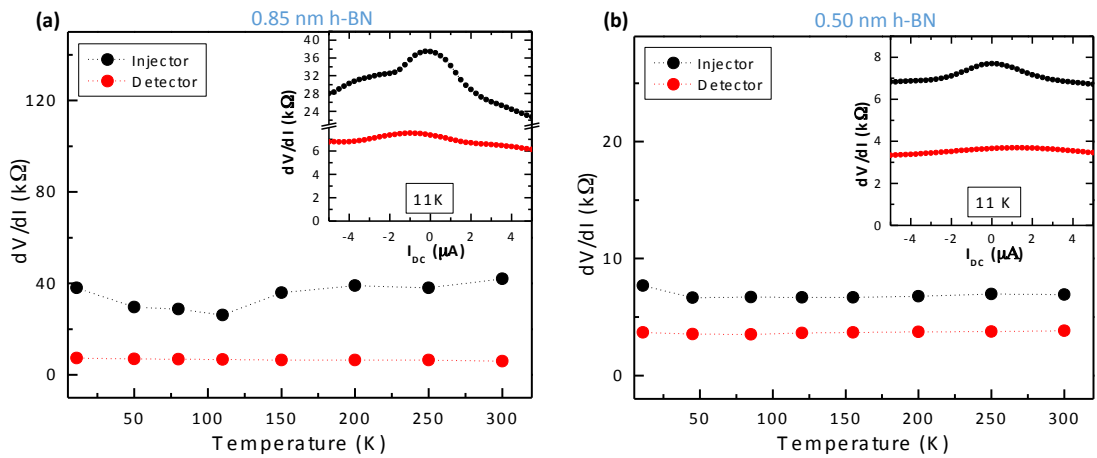

**Figure S1**: The temperature dependence of the zero bias contact resistance of the injector (black filled circles) and the detector (red filled circles) electrode. Inset: the differential contact resistance of the injector (black filled circles) and detector (red filled circles) electrodes as a function of bias current measured at 11 K. (a) Contact resistances for the 0.85 nm thick h-BN. (b) Contact resistances for the 0.50 nm thick h-BN tunnel barrier.

## II. Gate dependence of non-local MR signals

We show the gate dependence of the non-local MR signals measured using different thicknesses of the h-BN tunnel barrier. Figure S2a shows the MR signals, measured using 0.85 nm h-BN tunnel barrier, at room temperature for the applied back gate voltages ranging from -40 V to +30 V. The MR signal varies from 2 Ω to 5 Ω. The inset of Figure S2a shows the gate dependent resistance of the graphene channel where the Dirac point, maximum resistance, is at -30 V. Figure S2b shows the gate dependence of the MR signal measured using 0.50 nm thick h-BN tunnel barrier. The MR signals, which vary from 100 mΩ to 350 mΩ, are measured at 11 K. The graphene channel resistance as a function of gate voltage is shown in the inset to Figure S2b.

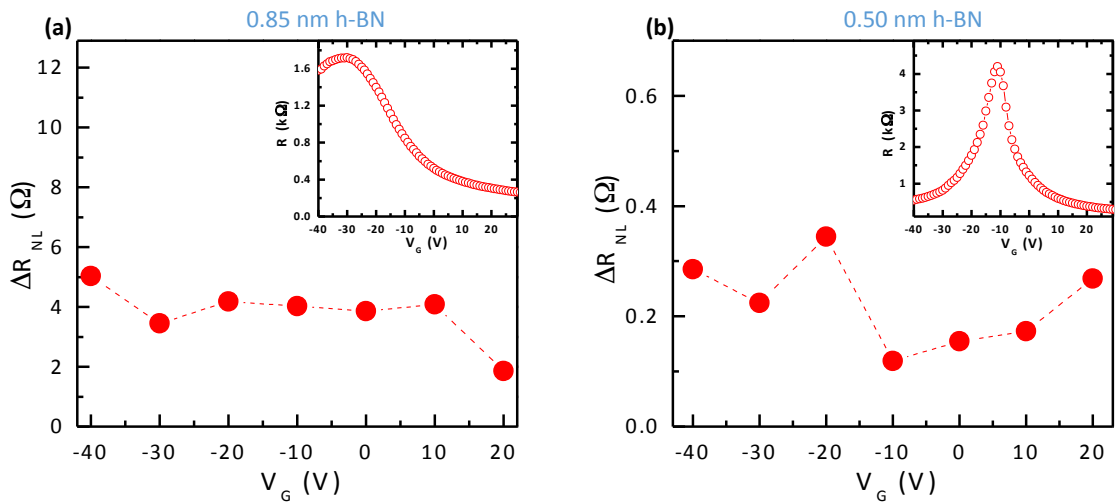

**Figure S2:** (a) The gate dependence of the magnitude of the non-local MR signal for 0.85 nm thick h-BN tunnel barrier at room temperature. Inset shows the resistance of the graphene channel as a function of applied back gate voltage. (b) Low temperature gate dependence of the magnitude of the MR signal using 0.50 nm thick h-BN tunnel barrier. Inset shows the graphene resistance vs applied gate voltage scan.

## III. Spin injection efficiency

The spin injection efficiency $\gamma$ is defined as the polarization of the injected current, $\gamma = \frac{I_{inj}^\uparrow - I_{inj}^\downarrow}{I_{inj}^\uparrow + I_{inj}^\downarrow}$. This is an important quantity that depends not only on the bulk and interfacial spin polarizations of the injector, but also on various quantities such as resistances, spin diffusion lengths, etc. It is valuable to understand the factors which determine the spin injection efficiency for h-BN tunnel barriers, especially considering that the spin lifetime depends strongly on the barrier thickness (Fig. 2 and 3 of main text). To this end, we derive the relation between $\gamma$ and the parameters extracted from the Hanle fit in the main text ($\sqrt{p_1 p_2}$, $\tau_s$, $\lambda$, $D$) and quantities determined experimentally (e.g. resistances, geometry, etc.). Because we are interested in the spin injection efficiency in zero applied field, we consider the Takahashi-Maekawa model[1] which is the zero field limit of the Aji model[2] used in the main text. The parameters $\tau_s$ (spin lifetime), $\lambda$ (spin diffusion length), and $D$ (diffusion constant) are used similarly in both models, but the polarizations are defined slightly differently. In the Takahashi-Maekawa model, the spin polarization of the ferromagnet is the asymmetry of the bulk conductivity $\left(p_{F,i} = \frac{\sigma_{F,i}^\uparrow - \sigma_{F,i}^\downarrow}{\sigma_{F,i}^\uparrow + \sigma_{F,i}^\downarrow}\right)$, where $i$ = 1 for injector and 2 for detector, and the spin polarization of the junction is the asymmetry of the interfacial conductance $\left(P_{J,i} = \frac{G_i^\uparrow - G_i^\downarrow}{G_i^\uparrow + G_i^\downarrow}\right)$. However, the Aji polarization $p_i$ is a weighted average of $p_{F,i}$ and $P_{J,i}$ given by $p_i = \frac{P_{J,i} R_i + p_{F,i} R_{F,i}}{R_{F,i} + R_i}$, where $R_i$ are the contact resistances and $R_{F,i} = \rho_{F,i} \lambda_{F,i}/A_i$ are the spin resistances of the injector ($i$=1) and detector ($i$=2) ferromagnets, $A_i$ are the junction areas and $\rho_{F,i}$ and $\lambda_{F,i}$ are the conductivity and spin diffusion lengths of the ferromagnets, respectively.

Using the approach of Takahashi and Maekawa[1], we calculate the spin injection efficiency to be

$$\gamma = \frac{2\left(\frac{p_{F,1} R_{F,1}}{(1-p_{F,1}^2)R_N} + \frac{P_{J,1} R_1}{(1-P_{J,1}^2)R_N}\right)\left(1 + \frac{2R_{F,2}}{(1-p_{F,2}^2)R_N} + \frac{2R_2}{(1-P_{J,2}^2)R_N}\right)}{\left(1 + \frac{2R_{F,1}}{(1-p_{F,1}^2)R_N} + \frac{2R_1}{(1-P_{J,1}^2)R_N}\right)\left(1 + \frac{2R_{F,2}}{(1-p_{F,2}^2)R_N} + \frac{2R_2}{(1-P_{J,2}^2)R_N}\right) - e^{-2L/\lambda}}$$  (S1)

where $L$ is the spacing between injector and detector, $R_N = R_{SQ}\frac{\lambda}{w}$ is the spin resistance of graphene, $w$ is the width of the graphene channel, and $R_{SQ}(=\frac{1}{\sigma})$ is the sheet resistance of graphene.

In the limit when contact resistance of the injector and detector electrodes are much larger than the spin resistance of the ferromagnet (i.e. $R_i \gg R_{F,i}$), the spin injection can be written as:

$$\gamma = \frac{2\left(\frac{P_{J,1} R_1}{(1-P_{J,1}^2)R_N}\right)\left(1 + \frac{2R_2}{(1-P_{J,2}^2)R_N}\right)}{\left(1 + \frac{2R_1}{(1-P_{J,1}^2)R_N}\right)\left(1 + \frac{2R_2}{(1-P_{J,2}^2)R_N}\right) - e^{-2L/\lambda}}$$  (S2)

In addition, the Aji polarization becomes equal to the junction polarization ($p_i = P_{J,i}$). This limit is satisfied for the h-BN barriers due to the low resistivity of the ferromagnet. Because this does not assume $R_i \gg R_N$, the spin injection efficiency could still be reduced by conductivity mismatch (between graphene and ferromagnet) if the barrier resistance is comparable to or lower than the graphene spin resistance.

For calculating $\gamma$, we use assume $P_{J,1} = p_1 = p_2 = \sqrt{p_1 p_2}$, which is a fit parameter from Eq. 2 in the main text. Using Eq. S2, we extract $\gamma = 0.022$ and $\gamma = 0.052$ in graphene for monolayer and multilayer h-BN tunnel barrier, respectively, for measurement conditions described in the main text (e.g. gate voltage, temperature, etc.). We note that calculated values of spin injection efficiencies are similar to the fitted values of $\sqrt{p_1 p_2}$ = 0.023 and $\sqrt{p_1 p_2}$ = 0.053 for monolayer and multilayer h-BN, respectively, from the main text. This illustrates that the spin injection efficiency is determined by the junction polarization and is not substantially reduced by the conductivity mismatch between the ferromagnet and the graphene channel. For both monolayer and multilayer h-BN tunnel barriers, we are in the limit where interfacial contact resistance is much higher than both the ferromagnetic spin resistance and the spin resistance of the graphene. An interesting question is whether the higher junction polarization for the thicker h-BN is due to spin filtering effects[3,4] predicted for h-BN tunnel barriers. To address this, further studies would be needed to determine if there is a systematic trend of enhanced spin injection efficiency with increased h-BN thickness.

References

1. S. Takahashi and S. Maekawa, Phys. Rev. B. **67**, 052409 (2003).

2. E. Sosenko, H. Wei, and V. Aji, Physical Review B **89**, 245436 (2014).

3. M. V. Kamalakar, A. Dankert, P. J. Kelly, and S. P. Dash, Scientific Reports **6**, 21168 (2016).

4. Q. Wu, L. Shen, Z. Bai, M. Zeng, M. Yang, Z. Huang, and Y. P. Feng, Physical Review Applied **2**, 044008 (2014).